\documentstyle[12pt]{article}

\begin{document}

\author{Hao Chen \\
Department of Mathematics\\
Zhongshan University\\
Guangzhou,Guangdong 510275\\
People's Republic of China}
\title{Schmidt numbers of low rank bipartite mixed states}
\date{October,2002}
\maketitle

\begin{abstract}
Schmidt numbers of bipartite mixed states ([1]) characterize the minimum Schmidt ranks of pure states that are needed to construct such mixed states. It is the minimum  number of degrees of freedom of a bipartite mixed state entangled between two parties. We give a lower bound of Schmidt numbers of  low rank bipartite mixed states and conclude that generic (i.e., all outside a measure zero set) low rank mixed states have relatively high Schmidt numbers and thus entangled. 
\end{abstract}

In quantum information theory the study of bipartite entanglement is of great importance. A bipartite pure state $|\psi>$ can always be described by its Schmidt decompsition; i.e., the representation of $|\psi>$ in an orthogonal product basis with minimum number of terms, $|\psi>=\Sigma_{i=1}^k p_i|a_i> \otimes|b_i>$ with positive reals $p_i$'s. The Schmidt rank $k$ is the number of nonvanishing terms in the representation. It is actually the rank of the reduced density matrix $\rho_B=Tr_B(|\psi><\psi|)$. The Schmidt ranks of pure states give a clear insight into the number of degrees of freedom that are entangled between two parties ([2]). A necessary condition for a pure state to be convertible by local quantum operations and classical communication(LOCC) to another pure state is that the Schmidt rank of the first pure state is larger than or equal to the Schmidt rank of the latter pure state; that is local operations and classical communication cannot increase the Schmidt rank of a pure state ([3]). The characterization of mixed state entanglement is a much harder task. A great effort has been devoted to detecting the presence of entanglement in a given mixed state (see [4],[5],[6],[7]) for the study of separability criteria. Terhal and P.Horodecki [1] defined the notation of  Schmidt numbers of  bipartite mixed states. For a bipartite mixed state $\rho$, it has Schmidt number $k$ if and only if for any decomposition $\rho=\Sigma_i p_i |v_i><v_i|$ for positive real
numbers $p_i$'s and pure states $|v_i>$'s, at least one of the pure states $%
|v_i>$'s has Schmidt rank at least $k$, and there exists such a
decomposition with all pure states $|v_i>$'s Schmidt rank at most $k$. Schmidt numbers of bipartite mixed states characterize the minimum Schmidt ranks of pure states that are needed to construct such mixed states. It is the minimum  number of degrees of freedom of a bipartite mixed state entangled between two parties. It is clear that the mixed states are entangled if their Schmidt numbers are bigger than 1. It is proved ([1]) that Schmidt number is entanglement monotone, ie., they cannot increase under local quantum operations and classical communication .
So we can naturally think Schmidt numbers of mixed states as a measure of their entanglement. \\

Because  Schmidt numbers of bipartite mixed states cannnot increase under LOCC,  it is desirable if people could compute Schmidt numbers of bipartite mixed states effectively and thus it would offer an effective-deciding criterion that two bipartite mixed states cannot be convertible by LOCC. However in this aspect Schmidt numbers are only calculated for very few bipartite mixed states, for example, it is calculated for "isotropic states" $\rho=\frac{1-F}{N^2-1}(I-|\Psi><\Psi|)+F|\Psi><|\Psi|$ on $H_A^N \otimes H_B^N$ , where $|\Psi>=\Sigma_{i=1}{N} |ii>$ in [1]. Some methods to relate Schmidt numbers of bipartite mixed states to "k-positive" maps and so-called "Schmidt number witness" have been developed in [1] and [8].\\

On the other hand, a method of  decomposing a mixed state by "edge" entangled states and separable state was developed in [9] [10]. From this method it is found that the low rank mixed state entanglement seems to be easier to understand and it is proved in [11] that for mixed states on $H_A^m \otimes H_B^n$ with ranks not larger than $max\{m,n\}$ PPT (positive partial transposition) is equivalent to separability.\\

In this paper we introduce linear subspaces $L_A(\rho)$ and $L_B(\rho)$ of $H_A^m$ and $H_B^n$ for  bipartite mixed states $\rho$ on $H_A^m \otimes H_B^n$, which are closely related to the Schmidt numbers of the mixed states $\rho$. Roughly speaking the smaller the dimensions of these linear subspaces are the biggger the Schmidt number of $\rho$ is (see Theorem 1). Then we give a lower bound for Schmidt numbers of low rank mixed states $\rho$ from $L_A(\rho)$ and $L_B(\rho)$.  On the other hand it is very easy to compute $L_A(\rho)$ or $L_B(\rho)$ from any representation of $\rho$ as a convex combination $\rho=\Sigma_i p_i |\psi_i><\psi_i|$ with $p_i$'s positive reals and pure states $|\psi_i>$. Thus we give an easy numerical method to give a lower bound for Schmidt numbers of  low rank bipartite mixed states. Sometimes the lower bound can be used to compute Schmidt numbers of mixed states exactly. Another implication of our result is that for generic rank $r < n$ mixed state on $H_A^m \otimes H_B^n$ (assume $m \leq n$ without loss of generality), the Schmidt numbers of these mixed states are at least $min\{\frac{n}{r},m\}$ and thus entangled. The results here also give an interpretation that low rank mixed state entanglement are easier to understand as in [11].\\

For any given bipartite mixed state $\rho$ on $H_A^m \otimes H_B^n$, $L_A(\rho)$(resp. $L_B(\rho))$ is the set of pure states $|a>$ in $H_A^m$ (resp. pure states $|b'>$ in $H_B^n$) such that $<a \otimes b |\rho|a \otimes b>=0$ for any pure state $|b>$ in $H_B^n$ (resp. $<a' \otimes b'|\rho|a' \otimes b'>=0$ for any pure state $|a'>$ in $H_A^m$). Since $<a \otimes b |(U_A \otimes U_B) \rho (U_A \otimes U_B)^{-1}|a\otimes b>= <(U_A^{-1}a)\otimes (U_B^{-1}b)|\rho| (U_A^{-1}a)\otimes (U_B^{-1}b)>$, $L_A((U_A\otimes U_B)\rho)$ is  $U_A^{-1} (L_A(\rho))$, i.e.,the  dimensions of $L_A(\rho)$ and $L_B(\rho)$ are invariant under local unitary operations.\\

 For a pure state $\rho=|\psi><\psi|$, from the above invariance under local unitary operations, we can compute $L_A(\rho)$ and $L_B(\rho)$ from the Schmidt decomposition of $|\psi>=\Sigma_{i=1}^k p_i|a_i>\otimes|b_i>$ with positive reals $p_i$, that is, $<a\otimes b|\rho|a \otimes b>=\Sigma_{i=1} p_i^2 |<a|a_i><b|b_i>|^2=0$ implies that $L_A(\rho)$ (resp. $L_B(\rho)$) is the orthogonal complementary in $H_A^m$ of the space spaned by pure states $|a_i>$'s (resp. $|b_i>$'s). Thus the Schmidt rank $k$ of the pure state $\rho$ is just the codimensions of the linear subspaeces of $L_A(\rho)$ and $L_B(\rho)$, that is, the linear subspaces we introduced as above can be thought as the degrees of freedom that are not entangled in the pure state. This point is manifested in the following result which asserts that the Schmidt number (i.e., the degrees of freedom entangled in two parties) of a bipartite mixed state $\rho$ is relatively high if the dimension of $L_A(\rho)$ or $L_B(\rho)$ is small.\\

{\bf Theorem 1.} {\em Let $\rho$ be a rank $r$ mixed state on $H_A^m \otimes H_B^n$ with Schmidt number $k$. Then $k \geq \frac{m-dim(L_A(\rho))}{r}$ and $k \geq \frac{n-dim(L_B(\rho))}{r}$.}\\

It is easy to see that for any bipartite mixed state $\rho$ on $H_A^m \otimes H_B^n$ ,\\$<a \otimes b |\rho|a \otimes b>=0$ is equivalent to $|a> \otimes |b>$ is in the kernel of $\rho$, thus it is equivalent to $|a> \otimes |b>$ is orthogonal to the range of $\rho$. This observation implies that $L_A(\rho)=\cap_{i=1}^r L_A(|v_i>)$ for pure states $|v_1>,...,|v_r>$ in $H_A^m \otimes H_B^n$ if they span the range of $\rho$.  We can now recall a lemma in [5], which asserts that for a mixed state of the form $\rho=\Sigma p_i |\phi_i><\phi_i|$, where $p_i$'s are positive reals and $|\phi_i>$'s are pure states, the range of $\rho$ is the linear span of pure states $|\phi_i>$'s. Combining these two observations together we can see that we can know some properties of Schmidt ranks of $|\phi_i>$'s for {\it any representation} of $\rho$ as $\Sigma p_i|\phi_i><\phi_i|$ from $L_A(\rho)$ or $L_B(\rho)$.\\

{\bf Proof of Theorem 1.} We just prove the coclusion $k \geq \frac{m-dim L_A(\rho)}{r}$. Another conclusion can be proved similarly. Take any  representation of the mixed state $\rho$ as $\rho= \Sigma_{i=1}^t p_i |v_i><v_i|$
with $p_i$'s are positive, and the maximal Schmidt rank of $|v_i>$'s is $k$. As
observed above, we only need to take $r$ linear independent vectors in 
$\{|v_1>,...,|v_t>\}$, say $|v_1>,....,|v_r>$ to compute $L_A(\rho)=\cap_{i=1}^{r}L_A(|v_i>)$. From our observation of pure states $dim L_A(|v_i>)=m-k(v_i)$, where $k(v_i)$ is the
Schmidt rank of $v_i$. Thus we know that $dim L_A(\rho) \geq m-\Sigma_{i=1}{r} k(v_i) \geq m-rk$,  since $k(v_i) \leq k$. The conclusion is proved.\\

For the convenience to use Theorem 1 we consider the coordinate form of the above linear subspaces $L_A(\rho)$ and $L_B(\rho)$. Let $\{|1>,...,|m>\}$ and $\{|1>,...,|n>\}$ be the standard orthogonal basis of $H_A^m$ and $H_B^n$ and $\{|11>,...,|1n>,....,|m1>,...,|mn>\}$ be the standard orthogonal base of  $H_A^m \otimes H_B^n$. We represent the matrix of $\rho$ in the base $\{|11>,...|1n>,...,|m1>,...,|mn>\}$, and consider $\rho$ as a blocked matrix $\rho=(\rho_{ij})_{1 \leq i \leq m, 1 \leq j \leq m}$ with each block $\rho_{ij}$ a $n \times n$ matrix corresponding to the $|i1>,...,|in>$ rows and the $|j1>,...,|jn>$ columns. For any pure state $|a>=r_1|1>+...+r_m|m> $in $H_A^m$,  the matrix of the Hermitian linear form $<a \otimes b |\rho| a \otimes b>$ of $|b>$ in $H_B^n$,  with respect to the the base $|1>,...,|n>$ ,is $\Sigma_{i,j} r_ir_j^{*} \rho_{ij}$. Let $\rho= \Sigma_{l=1}^{t} p_l |v_l><v_l|$ be any given representation of $\rho$ as a convex combination of projections
with $p_1,...,p_t >0$. Suppose $v_l=\Sigma_{i,j=1}^{m,n} a_{ijl} |ij>$ , $A=(a_{ijl})_{1\leq i \leq m, 1 \leq j \leq n, 1 \leq l \leq t}$ is the $mn \times t$ matrix. Then it is clear that the matrix representation of $\rho$ with the base $\{|11>,...,|1n>,...,|m1>,...,|mn>\}$ is $AP(A^{*})^{\tau}$, where $P$ is the diagonal matrix with diagonal entries $p_1,...,p_t$ (We always use $*$ denote the complex conjugate and $\tau$ indicate matrix transpositon). We may
consider the $mn\times t$ matrix $A$ as a $m\times 1$ blocked matrix with
each block $A_w$, where $w=1,...,m$, a $n\times t$ matrix corresponding to 
$\{|w1>,...,|wn>\}$. Then $\rho_{ij}=A_i P (A_j^{*})^{\tau}$ and thus $\Sigma_{ij}r_ir_j^{*} \rho_{ij}=\Sigma_{ij} r_ir_j^{*}A_i P (A_j^{*})^{\tau}=(\Sigma_i r_iA_i)P(\Sigma_j r_j^{*} A_j^{*})^{\tau}$. We note that $P$ is a strictly positive definate matrix , thus $L_A(\rho)$ is just the set of pure states $|a>=r_1|1>+...+r_m|m>$ in $H_A^m$ such that the matrix $\Sigma_i r_i A_i$ is the zero matrix (of size $n \times t)$. Similarly we can have the coordinate form of $L_B(\rho)$.\\

We can now return to the pure state case. Let $\rho=|v><v|$, $|v>=\Sigma_{i=1,j=1}^{m,n}a_{ij}$ be a pure state on $H_A^m \otimes H_B^n$. Consider $A=(a_{ij})_{1\leq i \leq m,1\leq j \leq n}$ be a $m \times n$ matrix, then $L_A(\rho)$ is just the pure states $r_1 |1>+...+r_m |m>$(in $H_A^m$)  such that $(r_1,...,r_m)A=0$ and $L_B(\rho)$ is just the pure states $r_1|1>+..+r_n|n>$ (in $H_B^n$) such that $A(r_1,...,r_n)^{\tau}=0$. They are of codimension $rank (A)$ (equal to the Schmidt rank of $|v>$ as well-known).\\

This can be easily generalized to mixed state case. Let $\rho=\Sigma_{k=1}^t p_i |v_k><v_k|$, where $p_i$ 's are positive, be a mixed state on $H_A^m \otimes H_B^n$, $|v_k>=\Sigma_{i=1,j=n}^{m,n} a_{ij}^k |ij>$ be the expansion in the standard base. We arrange the $t$ matrices of size $m \times n$ $A^k=(a_{ij}^k)_{1 \leq i \leq m, 1 \leq j \leq n}$,$k=1,...,t$ , in two different ways. $T_1=(A^1,...,A^t)$ is a matrix of size $m \times tn$ and $T_2=(A^1,...,A^t)^{\tau}$ is a matrix of size $tm \times n$ (note in each $A^k$ no transposition imposed). Then $L_A(\rho)$ is just the pure states $r_1 |1>+...+r_m |m>$(in $H_A^m$)  such that $(r_1,...,r_m)T_1=0$ and $L_B(\rho)$ is just the pure states $r_1|1>+..+r_n|n>$ (in $H_B^n$) such that $T_2(r_1,...,r_n)^{\tau}=0$. Theorem 1 can be re-read as the Schmidt number of $\rho$ is at least $\frac{rank(T_1)}{r}$ and $\frac{rank(T_2)}{r}$. From this point of view , Theorem 1 can be thought as a natural extension of the  previously well-known result that the Schmidt rank of $|v>$ is just the rank of the matrix $A$.\\

It follows immediately that\\

{\bf Proposition 1.} {\em Let $\rho=\Sigma_{k=1}^t p_k |v_k><v_k|$ be a rank $r < m$ mixed state on $H_A^m \otimes H_B^m$, where $p_1,...,p_t$ are
positive reals and $v_k=\Sigma_{ij} a_{ij}^k|ij>, A^k=(a_{ij}^k)_{1 \leq
i,j \leq m}$. Suppose that the linear span of all $rm$ rows of matrices $A^1,...,A^t$ is of dimension $m$. Then Schmidt number of $\rho$ is at least 
$\frac{m}{r}$,thus entangled.}\\

It is clear that the condition of Proposition 1 is satisfied by generic rank $r<m$ mixed states on $H_A^m \otimes H_B^m$.\\

Theorem 1 also implies that if a mixed state is mixed by not too many pure
states and one of these pure states has highest Schmidt rank, then the mixed
state has a relatively high Schmidt number. This can be sometimes used as an easy way to detect the entanglement in low rank mixed states.\\

{\bf Proposition 2.}{\em Let $\rho=\Sigma_{k=1}^t p_k |v_k><v_k|$ be a rank $r$ mixed
state on $H_A^m \otimes H_B^n$ with $r< m \leq n$, where $p_1,...,p_t$ are positive real
numbers and Schmidt rank of $v_1$ is $m$. Then Schmidt number of $\rho$ is
at least $\frac{m}{r}$ and thus entangled.}\\

Sometimes the lower bound in Theorem 1 can be used to give {\bf exact} results of Schmidt numbers for some bipartite mixed states.\\

{\bf Example 1.} Let $\rho=\Sigma_{k=1}^m p_i|\phi_k><\phi_k|$ be a mixed state on $H_A ^{mn} \otimes H_B^{mn}$ , where $p_1,...,p_m$ are positive reals and  $|\phi_k>=\Sigma_{i=1,j=1}^{n} a_{ij}^k |(k-1)m+i,(k-1)m+j>$ for $k=1,...,m$ . From this representation of $\rho$ we know that the Schmidt number of $\rho$ is at most $n$. Consider $A^k=(a_{ij}^k)_{1\leq i \leq n, 1\leq j \leq n}$ , $k=1,...,m$ , as matrices of size $n \times n$ , if $\Sigma_{k=1}^m rank(A^k) \geq mn-m+1$ then the Schmidt number of $\rho$ is exactly $n$. First we get $rank(T_1)=\Sigma_{k=1}^m rank(A^k) \geq mn-m+1$ and thus the Schmidt number of $\rho$ is at least $n-1+\frac{1}{m}$ from Theorem 1. On the other hand from the condition $\Sigma_{k=1}^m rank(A^k) \geq mn-m+1$, at least one of $A^k$, $k=1,...,m$, is of rank $n$ and thus the corresponding pure state is of Schmidt rank $n$. Thus we get the conclusion. We can observe the following case.\\

In the case  $m=3,n=4$, consider the following two mixed states $\rho_1=\frac{1}{4}\Sigma_{i=1}^4 |\phi_i><\phi_i|$ and $\rho_2=\frac{1}{3}\Sigma_{i=1}^3 |\psi_i><\psi_i|$ on $H_A^{12} \otimes H_B^{12}$, where,\\

$$
\begin{array}{cccccccccccc}
\phi_1= \frac{1}{\sqrt{3}}(|11>+|22>+|33>)\\
\phi_2= \frac{1}{\sqrt{2}}(|44>+|55>)\\
\phi_3= \frac{1}{\sqrt{3}}(|77>+|88>+|99>)\\
\phi_4= \frac{1}{\sqrt{2}}(|11,11>+|12,12>)\\
\psi_1= \frac{1}{2}(|11>+|22>+|33>+|44>)\\
\psi_2= \frac{1}{\sqrt{3}}(|55>+|66>+|77>)\\
\psi_3= \frac{1}{\sqrt{3}}(|10,10>+|11,11>+|12,12>)
\end{array}
$$

Thus we know that the Schmidt number of $\rho_1$ is 3 and the Schmidt number of $\rho_2$ is 4. \\

{\bf Example 2.} Let $\rho=\Sigma_{i=1}^r p_i |\phi_i><\phi_i|$, where $p_1,...,p_r$ are positive reals, be a rank $r$ mixed state on $H_A^m \otimes H_B^n$ with $rm \leq n$. Suppose the matrix $T_2$ (size $rm \times n$ ) of the above representation of $\rho$ is of full rank $rm$. From Theorem 1 the Schmidt number of $\rho$ is at least  $\frac{rank(T_2)}{r}=m$. Since $m$ is the highest possible value of Schmidt numbers, the Schmidt number of $\rho$ is exactly $m$.\\

Let $\rho=\Sigma_{i=1}^3 p_i |\phi_i><\phi_i|$, where $p_1,p_2,p_3$ are positive reals, and \\

$$
\begin{array}{ccccccccccccc}
\phi_1=\frac{1}{3}(|11>+|12>+|13>+|15>+|17>+|22>+|28>+|33>+|39>)\\
\phi_2=\frac{1}{\sqrt{7}}(|14>+|15>+|16>+|25>+|27>+|36>+|39>)\\
\phi_3=\frac{1}{2}(|17>+|19>+|28>+|39>)
\end{array}
$$

be a rank $3$ mixed state on $H_A^3 \otimes H_B^9$. We can check that its $T_2$ is a rank 9 matrix (size $9 \times 9$). Thus the Schmidt number of $\rho$ is 3.\\ 

For a given mixed state $\rho=\Sigma_i p_i |v_i><v_i|$ with positive reals $p_i$'s, it is clear from the definition of Schmidt number we know that the Schmidt number of $\rho$ is at most the maximal of Schmidt ranks of $|v_i>$'s. On the other hand there is a lower bound of Schmidt numbers for bipartite mixed states from Theorem 1. Hence in some cases we can compare Schmidt numbers from the above fact and know the first one cannot be convertible to the latter mixed state by local operations and classical communication.\\

{\bf Example 3.} Let $\rho=\Sigma_{i=1}^{m-2} p_i |\phi_i><\phi_i|$ , where $p_i$'s are positive reals and $|\phi_i>=\frac{1}{\sqrt{3}}(|ii>+|i+1,i+1>+|i+2,i+2>)$ for $i=1,...,m-2$ , be a mixed state on $H_A^m \otimes H_B^m$. It is clear that the Schmidt number of $\rho$ is at most $3$.\\

Let $\rho'=\frac{1}{2}(|\psi_1><\psi_1|+|\psi_2><\psi_2|)$ be on a mixed state on $H_A^n \otimes H_B^n$ with $n \geq 8$, where $|\psi_1>,|\psi_2>$ are pure states in $H_A^n \otimes H_B^n$. Suppose the Schmidt number of $|\psi_1>$ or $|\psi_2>$ is $n$. From Proposition 2, the Schmidt number of $\rho'$ is at least $\frac{n}{2} \geq 4$. Thus $\rho$ cannot be convertible to $\rho'$ by local operations and classical communication.\\

Because the above discussion about the coordinate form of Theorem 1, we can easily calculate the lower bound of Schmidt numbers of low rank bipartite mixed states. This gives us the following conclusion.\\

{\bf Theorem 2.} {\em Generic rank $r<n$ mixed states on $H_A^m \otimes H_B^n$ (assume $m \leq n$ without loss of generality) have their Schmidt numbers at least $min\{\frac{n}{r},m\}$ and thus entangled.} \\

We consider the spectral decomposition $\rho=\Sigma_{k=1}^r p_i |v_k><v_k|$ of rank $r$ mixed states on $H_A^m \otimes H_B^n$, where $p_k$ and $|v_k>$,$k=1,...,r$ , is eigenvalue and corresponding eigenvector. Let $|v_k>=\Sigma_{ij} a_{ij}^k |ij>$ be the expansion in the standard base. Consider $A^k=(a_{ij}^k)_{1\leq i \leq m , 1\leq j \leq n}$ as a $m \times n$ matrix for $k=1,...,r$. We know that the only condition imposed on the matrices $A^1,...,A^r$ is that they are orthogonal when they are considered as vectors in $H_A^m \otimes H_B^n$. Under this condition it is clear that the matrix $T_2=(A^1,...,A^r)^{\tau}$ (of size $ rm \times n$)can reach the highest possible rank $min \{rm,n\}$, since we can take rows of $T_2$ to be the $min \{rm,n\}$  orthogonal vectors in $C^n$. In this case the Schmidt number of the corresponding $\rho$ is at least $min \{\frac{n}{r},m\}$ from Theorem 1 and its coordinate form. In the space of all possible matrices $T_2$'s corresponding spectral decomposition of rank $r$ mixed states on $H_A^m \otimes H_B^n$, outside the set defined by the condition $rank (T_2)< min \{rm,n\}$, it is known that the Schmidt numbers of corresponding mixed states are at least $min\{ \frac{n}{r},m\}$. On the other hand, the condition $rank(T_2)< min\{rm,n\}$ is equivalent to some algebraic equations on variables $a_{ij}^k$'s , thus the set defined by this condition is a measure zero set. Correspondingly we know that generic rank $r<n$ mixed states on $H_A^m \otimes H_B^n$ have their Schmidt numbers at least $ min\{\frac{n}{r},m\}$.\\

In conclusion, we proved a lower bound for Schmidt numbers of bipartite mixed states. This lower bound can be applied easily to low rank bipartite mixed states. From this lower bound it is known that generic low rank bipartite mixed states have relatively high Schmidt numbers and thus entangled. We can compute Schmidt numbers exactly for some mixed states by this lower bound as shown in Examples. This lower bound can also be used effectively to determine that some mixed states cannot be convertible to other mixed states by local operations and classical communication.\\

The author acknowledges the support from NNSF China, Information Science
Division, grant 69972049.\\

e-mail: chenhao1964cn@yahoo.com.cn\newline

\begin{center}
REFERENCES
\end{center}

1.B.M.Terhal and P.Horodecki, Phys. Rev. A R040301, 61(2000)\\

2.A.Peres, Quantum Theory; Concepts and Methods, (Kluwer Academic Publisher, Dordrecht 1993)\\

3.H.-K.Lo and S.Popescu, quant-ph/9707038\\

4.A.Peres, Phys. Rev. Lett., 77, 1413(1996)\\

5.P.Horodecki, Phys Lett A 232 (1997)\\

6.C.H.Bennett, D.P.DiVincenzo, T.Mor,P.W.Shor, J.A.Smolin and T.M. Terhal, Phys Rev. Lett. 82, 5385 (1999)\\

7.A.C.Doherty, P.A.Parrilo and F.M.Spedalier, Phys. Rev. Lett., 88, 187904(2002)\\

8.A.Sanpera,D.Bruss and M.Lewenstein, Phys.Rev. A 63, 050301(R)(2001)\\

9.M.Lewenstein and A.Sanpera, Phys.Rev.Lett. 80, 2261(1998)\\

10.A.Sanpera,R,Tarrach and G.Vidal, Phys.Rev. A 56, 826(1998)\\

11.P.Horodecki, M.Lewenstein, G.Vidal and I.Cirac, Phys. Rev. A 62, 032302 (2000)\\

\end{document}